\documentclass[10pt]{article}
\usepackage{graphicx}
\usepackage{amsmath}
\usepackage{amssymb}
\usepackage{caption2}
\begin{document}
\bibliographystyle{prsty}
\begin{center}
{\large {\bf \sc{  Does vacuum saturation work for the higher dimensional vacuum condensates in the QCD sum rules?   }}} \\[2mm]
Zhi-Gang  Wang \footnote{E-mail: zgwang@aliyun.com.  }     \\
 Department of Physics, North China Electric Power University, Baoding 071003, P. R. China
\end{center}

\begin{abstract}
In the QCD sum rules for the tetraquark (molecular) states, the higher dimensional vacuum condensates play an important role in extracting the tetraquark masses. We carry out the operator product expansion up to  the vacuum condensates of dimension-10  and observe that the vacuum condensates of dimensions $6$, $8$ and $10$ have the same expressions but opposite signs for the $C\gamma_5\otimes \gamma_\mu C$-type  and  $C\otimes \gamma_\mu C$ type four-quark currents, which make their influences distinguishable, and they are excellent channels to examine the vacuum saturation approximation. We introduce a parameter $\kappa$ to parameterize the derivation from the vacuum saturation or factorization approximation, and choose two sets parameters to examine the influences on the predicted tetraquark masses, which can be confronted to the experimental data in the future.  In all the channels,  smaller value of the $\kappa$ leads to better convergent behavior in the operator product expansion, which favors the vacuum saturation approximation.
\end{abstract}



\section{Introduction}

 In the QCD sum rules, we construct the quark (or quark-gluon or gluon) currents to interpolate the hadrons with the same quantum numbers, then introduce the two-point correlation functions and accomplish the operator product expansion in the deep Euclidean space, the perturbative contributions are embodied in the Wilson's coefficients, while the nonperturbative contributions are absorbed in the vacuum condensates.   We have to resort to non-zero vacuum expectation values for  the normal-ordered   quark-gluon operators to describe the hadron properties in a satisfactory way.

 Except for the quark condensates $\langle\bar{q}q\rangle$ with $q=u$, $d$, which indicate  spontaneous breaking of the Chiral symmetry through the Gell-Mann-Oakes-Renner relation
 $f^2_{\pi}m^2_{\pi}=-2(m_u+m_d)\langle\bar{q} q\rangle$ \cite{GMOR}, where the $f_\pi$ is the decay constant of the pion,  other vacuum condensates, such as the $\langle\bar{q} g_s\sigma G q\rangle$,
 $\langle\bar{q} q\bar{q} q\rangle$, $\cdots$ are just parameters introduced by hand to describe the nonperturbative QCD vacuum,
 we can parameterize the non-perturbative properties  in one way or the other.

There are a large number of four-quark condensates which can have a nonzero vacuum expectation, such as the $\langle \bar{q} q\bar{q}q\rangle$, $\langle \bar{q} i\gamma_5 q\bar{q}i\gamma_5q\rangle$, $\langle \bar{q} \gamma_{\mu}q\bar{q}\gamma^{\mu}q\rangle$,
$\langle \bar{q} \gamma_{\mu}t^a q\bar{q}\gamma^{\mu}t^aq\rangle$, $\langle \bar{q} \gamma_{\mu}\gamma_5t^a q\bar{q}\gamma^{\mu}\gamma_5t^aq\rangle$, etc, where $t^a=\frac{\lambda^a}{2}$, the $\lambda^a$ is the Gell-Mann matrix.
Up to now, none of those four-quark vacuum condensates are well constrained or understood in any meaningful way. The commonly adopted  method
 is the so-called vacuum saturation (or factorization) approximation, we  insert a complete set of intermediate  states between two $\bar{q}$ and $q$ quarks
and  assume  the vacuum contribution dominates the sum of all the intermediate  states, and obtain
 the squared Chiral condensate  $\langle0|:\bar{q} q:|0\rangle^2$,
\begin{eqnarray}
 \langle0|:\bar{q}_{\alpha}^{i} q_{\beta}^{j}\bar{q}_{\lambda}^{m} q_{\tau}^{n}:|0\rangle
  &=&\frac{1}{16N_c^2}\langle0|:\bar{q} q:|0\rangle^2
 \left(\delta_{ij}\delta_{mn}\delta_{\alpha\beta}\delta_{\lambda\tau}-\delta_{in}\delta_{jm}\delta_{\alpha\tau}\delta_{\beta\lambda} \right)\, ,
  \end{eqnarray}
 where the $i$, $j$, $m$ and $n$ are color indexes, the $\alpha$, $\beta$, $\lambda$ and $\tau$ are Dirac spinor indexes.
 In the original works, Shifman,  Vainshtein and  Zakharov took
 the vacuum saturation hypothesis according to  two reasons \cite{SVZ79-1,SVZ79-2},  one reason is the rather large value of the quark (or Chiral) condensate $\langle\bar{q}q\rangle$, the other is the duality between the quark and physical states, which implies that counting both the quark and
physical states may well become a double counting because  they reproduce each other \cite{SVZ79-1,SVZ79-2}. In additional, according to the arguments of Shifman,  Vainshtein and  Zakharov and Ioffe \cite{SVZ79-1,SVZ79-2,Ioffe-SB}, the accuracy of factorization hypothesis is of order
$\frac{1}{N_c^2}$, the vacuum saturation (or factorization) works well in the large $N_c$ limit $\frac{1}{N_c^2}\sim 0$ \cite{Novikov--shifman}, in reality,
$N_c = 3$, $\frac{1}{N_c^2}\sim 10\%$.

We can introduce a parameter $\kappa$ to take into account the violation of the vacuum saturation (or factorization) approximation \cite{Narison-kappa},
\begin{eqnarray}
\langle0|:\bar{q} q:|0\rangle^2 &\to& \kappa\, \langle0|:\bar{q} q:|0\rangle^2\, ,
\end{eqnarray}
the value $\kappa=1$ stands for the vacuum saturation approximation, while the value $\kappa>1$ parameterizes its violation. A number of values have been
obtained, which depend on the studied channels. The existing estimations range from $\kappa=1$  \cite{Ioffe-Book} to $2\sim3$ \cite{Narison-Book} and even up to  6 \cite{Review-kappa-kappa}.

In the QCD sum rules for the traditional mesons, we usually carry out the operator product expansion up to the vacuum condensates of dimension $6$,
 the four-quark condensate $\langle\bar{q}q\rangle^2$ is always companied with the strong fine-structure constant $\alpha_s(\mu)$, the net effect is greatly depressed,
  and plays a minor important role,
the deviation from the value $\kappa=1$ cannot make much difference, although the value $\kappa>1$ can lead to better QCD sum rules in some cases \cite{Review-kappa-kappa}.
While in the QCD sum rules for the hidden-charm (or hidden-bottom) tetraquark (molecular) states, we usually carry out the operator product expansion up to the vacuum condensates of dimension $10$, the vacuum condensate $\langle\bar{q}q\rangle^2$   plays a very important role, the derivations from $\kappa=1$ make much diffidence on the predictions \cite{WZG-CTP-DvDvDv}.

 In the present work, we choose the hidden-charm axialvector and vector four-quark currents as an example to examine the validity of the vacuum saturation (or factorization) approximation in details.

The article is arranged in the form:  we obtain the QCD sum rules for the  axialvector and vector tetraquark  states  in section 2; in section 3, we present the numerical results and discussions; section 4 is reserved for our conclusion.

\section{QCD sum rules for  the hidden-charm   tetraquark  states }
Let us write down  the  correlation functions $\Pi_{\mu\nu}(p)$  in the QCD sum rules firstly,
\begin{eqnarray}
\Pi_{\mu\nu}(p)&=&i\int d^4x e^{ip \cdot x} \langle0|T\left\{J_\mu(x)J_\nu^{\dagger}(0)\right\}|0\rangle \, ,
\end{eqnarray}
where  $J_\mu(x)=J_\mu^A(x)$ and $J_\mu^V(x)$,
\begin{eqnarray}
J_\mu^A(x)&=&\varepsilon^{ijk}\varepsilon^{imn}u^{Tj}(x)C\gamma_5c^k(x) \bar{d}^m(x)\gamma_\mu C \bar{c}^{Tn}(x)\, ,\nonumber\\
J_\mu^V(x)&=&\varepsilon^{ijk}\varepsilon^{imn}u^{Tj}(x)Cc^k(x) \bar{d}^m(x)\gamma_\mu C \bar{c}^{Tn}(x)\, ,
\end{eqnarray}
where the $i$, $j$, $k$, $m$ and $n$ are color indexes, the charge conjugation matrix $C=i\gamma^2\gamma^0$.
The simple four-quark currents $J_\mu^A(x)$ and $J_\mu^V(x)$, which lead to simple analytical expressions of the QCD sum rules,  have both negative and positive charge-conjugation components. Now we write down the two components explicitly,
$\sqrt{2}J_\mu^A(x)=J^{SA}_{-,\mu}(x)+J^{SA}_{+,\mu}(x)$,
$\sqrt{2}J_\mu^V(x)= J^{PA}_{-,\mu}(x)+J^{PA}_{+,\mu}(x)$,
where
\begin{eqnarray}
J^{SA}_{-,\mu}(x)&=&\frac{\varepsilon^{ijk}\varepsilon^{imn}}{\sqrt{2}}\Big[u^{Tj}(x)C\gamma_5c^k(x) \bar{d}^m(x)\gamma_\mu C \bar{c}^{Tn}(x)-u^{Tj}(x)C\gamma_\mu c^k(x)\bar{d}^m(x)\gamma_5C \bar{c}^{Tn}(x) \Big] \, ,\nonumber\\
J^{SA}_{+,\mu}(x)&=&\frac{\varepsilon^{ijk}\varepsilon^{imn}}{\sqrt{2}}\Big[u^{Tj}(x)C\gamma_5c^k(x) \bar{d}^m(x)\gamma_\mu C \bar{c}^{Tn}(x)+u^{Tj}(x)C\gamma_\mu c^k(x)\bar{d}^m(x)\gamma_5C \bar{c}^{Tn}(x) \Big] \, , \nonumber \\
\end{eqnarray}
\begin{eqnarray}
J^{PA}_{-,\mu}(x)&=&\frac{\varepsilon^{ijk}\varepsilon^{imn}}{\sqrt{2}}\Big[u^{Tj}(x)Cc^k(x) \bar{d}^m(x)\gamma_\mu C \bar{c}^{Tn}(x)-u^{Tj}(x)C\gamma_\mu c^k(x)\bar{d}^m(x)C \bar{c}^{Tn}(x) \Big] \, ,\nonumber\\
J^{PA}_{+,\mu}(x)&=&\frac{\varepsilon^{ijk}\varepsilon^{imn}}{\sqrt{2}}\Big[u^{Tj}(x)Cc^k(x) \bar{d}^m(x)\gamma_\mu C \bar{c}^{Tn}(x)+u^{Tj}(x)C\gamma_\mu c^k(x)\bar{d}^m(x)C \bar{c}^{Tn}(x) \Big] \, ,
\end{eqnarray}
 and the superscripts $S$, $A$ and $P$ represent the scalar, axialvector and pseudoscalar (anti)diquark operators, respectively. The axialvector currents $J^{SA}_{+,\mu}(x)$ and $J^{SA}_{-,\mu}(x)$ couple potentially to the hidden-charm tetraquark states with the $J^{PC}=1^{++}$ and $1^{+-}$ respectively which have almost degenerated  masses but slightly different pole residues \cite{WZG-HT-PRD,WZG-PRD-hidden-charm}, while the vector currents $J^{PA}_{+,\mu}(x)$ and $J^{PA}_{-,\mu}(x)$ couple potentially to the hidden-charm tetraquark states with the $J^{PC}=1^{-+}$ and $1^{--}$ respectively which also have almost degenerated  masses
but slightly different pole residues \cite{Wang-tetra-formula,WZG-4260-4360-4660}. In the present work, we will not distinguish the charge-conjugation, and choose the simple four-quark currents $J_\mu^A(x)$ and $J_\mu^V(x)$, which couple potentially to the axialvector and vector hidden-charm tetraquark states with the $J^P=1^+$ and $1^-$, respectively, as we are only interested in the tetraquark masses.

The axialvector current $J^A_\mu(x)$ is of the $C\gamma_5\otimes \gamma_\mu C $ type, while the vector current $J^V_\mu(x)$ is of the $C\otimes \gamma_\mu C $ type,
the relative P-wave is tacitly or implicitly embodied in the pseudoscalar diquark. On the other hand, we can construct the $C\gamma_5\otimes \gamma_5\gamma_\mu C $ type vector current, where the relative P-wave is tacitly or implicitly embodied in the vector diquark \cite{WZG-4260-4360-4660}, or introduce a relative P-wave between the diquark and antidiquark  explicitly, and construct the
\begin{eqnarray}
&&C\gamma_5 \otimes \stackrel{\leftrightarrow}{\partial}_\mu\otimes\gamma_5 C \, ,\nonumber\\
&&C\gamma_\alpha \otimes \stackrel{\leftrightarrow}{\partial}_\mu\otimes\gamma^\alpha C \, ,\nonumber\\
&&C\gamma_\mu \otimes \stackrel{\leftrightarrow}{\partial}_\alpha\otimes\gamma^\alpha C +C\gamma^\alpha \otimes \stackrel{\leftrightarrow}{\partial}_\alpha\otimes\gamma_\mu C \, ,\nonumber\\
&&C\gamma_5 \otimes \stackrel{\leftrightarrow}{\partial}_\mu\otimes\gamma_\nu C+C\gamma_\nu \otimes \stackrel{\leftrightarrow}{\partial}_\mu\otimes\gamma_5 C-C\gamma_5 \otimes \stackrel{\leftrightarrow}{\partial}_\nu\otimes\gamma_\mu C-C\gamma_\mu \otimes \stackrel{\leftrightarrow}{\partial}_\nu\otimes\gamma_5 C \, ,
\end{eqnarray}
type vector currents, where $\stackrel{\leftrightarrow}{\partial}_\mu=\overrightarrow{\partial}_\mu-\overleftarrow{\partial}_\mu$ \cite{WZG-Y4260-vector-1,WZG-Y4260-vector-2}. As the analytical expressions of the QCD spectral densities for those currents are lengthy and do not have simple relations with  that for the $C\gamma_5\otimes \gamma_\mu C $ type current. We prefer the currents $J^A_\mu(x)$ and $J^V_\mu(x)$, because they lead to the QCD spectral densities which have simple relations with each other.

At the hadron side, we  isolate the contributions of the ground state hidden-charm  tetraquark   states $Z_c$,
\begin{eqnarray}
\Pi_{\mu\nu}(p)&=&\frac{\lambda_{Z}^2}{M_{Z}^2-p^2}\left(-g_{\mu\nu} +\frac{p_\mu p_\nu}{p^2}\right) +\cdots \, \, , \nonumber\\
               &=&\Pi(p^2)\left(-g_{\mu\nu} +\frac{p_\mu p_\nu}{p^2}\right) +\cdots \, \, ,
\end{eqnarray}
where the $M_Z$ and $\lambda_Z$ are masses and pole residues of the hidden-charm  tetraquark states $Z_c$, respectively,  the $\lambda_{Z}$ are defined by $ \langle 0|J_\mu(0)|Z_c(p)\rangle=\lambda_{Z} \,\varepsilon_\mu$,
the $\varepsilon_\mu$ are the polarization vectors of the axialvector and vector tetraquark states. We resort to the component $\Pi(p^2)$ to investigate  the axialvector and vector tetraquark states.

 We accomplish  the operator product expansion up to the vacuum condensates  of dimension-10 consistently, which are the vacuum expectations  of the quark-gluon operators of the order $\mathcal{O}(\alpha_s^k)$ with $k\leq1$.  In calculations, we assume vacuum saturation for the  higher dimensional  vacuum condensates and factorize the higher dimensional  vacuum condensates into the product of the lower ones.   Then we write the correlation functions $\Pi(p^2)$ in the form,
 \begin{eqnarray}
 \Pi(p^2)&=&\int_{4m_c^2}^{\infty} ds \frac{\rho_{QCD}(s)}{s-p^2}\, ,
 \end{eqnarray}
 through dispersion relation, where the $\rho_{QCD}$ are the spectral densities at the quark-gluon level.

 We  implement the quark-hadron duality below the continuum thresholds $s_0$ and accomplish Borel transform  with respect to
the variable $P^2=-p^2$ to obtain  the  QCD sum rules,
\begin{eqnarray}\label{QCDSR}
\lambda^2_{Z}\, \exp\left(-\frac{M^2_{Z}}{T^2}\right)= \int_{4m_c^2}^{s_0} ds\, \rho_{QCD}(s) \, \exp\left(-\frac{s}{T^2}\right) \, ,
\end{eqnarray}
where the $T^2$ are the Borel parameters,  $\rho_{QCD}(s)=\rho_A(s)$ and $\rho_V(s)$, $\rho_A(s)=\sum\limits_{i}\rho^A_{i}(s)$, $\rho_V(s)=\sum\limits_{i}\rho^V_{i}(s)$ with $i=0$, $3$, $4$, $5$, $6$, $7$, $8$ and $10$,
 \begin{eqnarray}
\rho^A_0(s)&=& \frac{1}{1024\pi^6} \int dydz
 \,{y}{z}(1-y-z)^2 \left(s-\overline{m}_c^2\right)^3(5s-\bar{m}_c^2) \, ,
\end{eqnarray}

\begin{eqnarray}
\rho^A_3(s)&=& -\frac{{m}_c\langle\bar{q}q\rangle}{32\pi^4} \int dydz
 \,{y}(1-y-z) \left(s-\overline{m}_c^2\right)(7s-3\overline{m}_c^2) \, ,
\end{eqnarray}

\begin{eqnarray}
\rho^A_4(s)&=& -\frac{ {m}_c^2}{384\pi^4}\langle\frac{\alpha_{s}{G}{G}}{\pi}\rangle \int dydz\,\frac{z(1-y-z)^2}{y^2}(2s-\overline{m}_c^2)  \nonumber\\
&& +\frac{ 1}{384\pi^4}\langle\frac{\alpha_{s}{G}{G}}{\pi}\rangle \int dydz\,z(1-y-z)s(s-\overline{m}_c^2)\, ,
\end{eqnarray}

\begin{eqnarray}
\rho^A_5(s)&=& \frac{{m}_c\langle\bar{q}g_{s}\sigma Gq\rangle}{64\pi^4} \int dydz
 \,y(5s-3\overline{m}_c^2) \nonumber\\
&& -\frac{{m}_c\langle\bar{q}g_{s}\sigma Gq\rangle}{64\pi^4} \int dydz
 \,\frac{z(1-y-z)}{y}(2s-\overline{m}_c^2)  \, ,
\end{eqnarray}

\begin{eqnarray}
\rho^A_6(s)&=& \frac{{m}_c^2\langle\bar{q}q\rangle^2}{12\pi^2} \int dy  \, ,
\end{eqnarray}

\begin{eqnarray}
\rho^A_7(s)&=&\frac{{m}_c^3\langle\bar{q}q\rangle}{288\pi^2}\langle\frac{\alpha_{s}{G}{G}}{\pi}\rangle \int dydz
 \, \frac{(1-y-z)(y+z)}{y^3}\left(1+\frac{2s}{T^2}\right) \delta\left(s-\overline{m}_c^2\right) \nonumber\\
&& -\frac{{m}_c\langle\bar{q}q\rangle}{96\pi^2}\langle\frac{\alpha_{s}{G}{G}}{\pi}\rangle  \int dydz
 \, \frac{z(1-y-z)}{y^2}\left[3+2s\,\delta\left(s-\overline{m}_c^2\right)\right]\nonumber\\
&& -\frac{{m}_c\langle\bar{q}q\rangle}{576\pi^2}\langle\frac{\alpha_{s}{G}{G}}{\pi}\rangle  \int dydz
 \,\left[3+2s\,\delta\left(s-\overline{m}_c^2\right)\right]\nonumber\\
 && -\frac{{m}_c\langle\bar{q}q\rangle}{576\pi^2}\langle\frac{\alpha_{s}{G}{G}}{\pi}\rangle  \int dy
 \,y\left[3+2s\,\delta\left(s-\widetilde{m}_c^2\right)\right] \, ,
\end{eqnarray}

\begin{eqnarray}
\rho^A_8(s)&=&-\frac{{m}_c^2\langle\bar{q}q\rangle\langle\bar{q}g_{s}\sigma Gq\rangle}{24\pi^2}\int dy
  \,\left(1+\frac{s}{T^2}\right) \delta\left(s-\widetilde{m}_c^2\right) \nonumber\\
&& +\frac{{m}_c^2\langle\bar{q}q\rangle\langle\bar{q}g_{s}\sigma Gq\rangle}{48\pi^2}\rangle  \int dy
 \,\frac{1}{y}\delta\left(s-\widetilde{m}_c^2\right) \, ,
\end{eqnarray}

\begin{eqnarray}
\rho^A_{10}(s)&=&\frac{{m}_c^2\langle\bar{q}g_{s}\sigma Gq\rangle^2}{192\pi^2T^6}\int dy  \,s^2\, \delta\left(s-\widetilde{m}_c^2\right) \nonumber\\
&&-\frac{{m}_c^2\langle\bar{q}g_{s}\sigma Gq\rangle^2}{192\pi^2T^4}\int dy
  \,\frac{s}{y} \delta\left(s-\widetilde{m}_c^2\right) \nonumber\\
&& -\frac{{m}_c^4\langle\bar{q}q\rangle^2}{108T^4}\langle\frac{\alpha_{s}{G}{G}}{\pi}\rangle   \int dy
 \,\frac{1}{y^3} \delta\left(s-\widetilde{m}_c^2\right) \nonumber\\
&& +\frac{{m}_c^2\langle\bar{q}q\rangle^2}{36T^2}\langle\frac{\alpha_{s}{G}{G}}{\pi}\rangle  \int dy
 \,\frac{1}{y^2} \delta\left(s-\widetilde{m}_c^2\right) \nonumber\\
&& +\frac{{m}_c^2\langle\bar{q}q\rangle^2}{216T^6}\langle\frac{\alpha_{s}{G}{G}}{\pi}\rangle   \int dy
 \, s^2\, \delta\left(s-\widetilde{m}_c^2\right) \, ,
\end{eqnarray}

\begin{eqnarray}
\rho^V_0(s)&=& \rho^A_0(s) \, ,
\end{eqnarray}

\begin{eqnarray}
\rho^V_3(s)&=& -\frac{{m}_c\langle\bar{q}q\rangle}{32\pi^4} \int dydz
 \,{y}(1-y-z) \left(s-\overline{m}_c^2\right)^2 \, ,
\end{eqnarray}

\begin{eqnarray}
\rho^V_4(s)&=& \rho^A_4(s)\, ,
\end{eqnarray}

\begin{eqnarray}
\rho^V_5(s)&=& \frac{{m}_c\langle\bar{q}g_{s}\sigma Gq\rangle}{64\pi^4} \int dydz
 \,y(s-\overline{m}_c^2) \nonumber\\
&& +\frac{{m}_c\langle\bar{q}g_{s}\sigma Gq\rangle}{64\pi^4} \int dydz
 \,\frac{z(1-y-z)}{y}(2s-\overline{m}_c^2)  \, ,
\end{eqnarray}

\begin{eqnarray}
\rho^V_6(s)&=& -\rho^A_6(s)  \, ,
\end{eqnarray}

\begin{eqnarray}
\rho^V_7(s)&=&\frac{{m}_c^3\langle\bar{q}q\rangle}{288\pi^2}\langle\frac{\alpha_{s}{G}{G}}{\pi}\rangle \int dydz
 \, \frac{(1-y-z)(y+z)}{y^3} \delta\left(s-\overline{m}_c^2\right) \nonumber\\
&& -\frac{{m}_c\langle\bar{q}q\rangle}{96\pi^2}\langle\frac{\alpha_{s}{G}{G}}{\pi}\rangle  \int dydz
 \, \frac{z(1-y-z)}{y^2}\nonumber\\
&& -\frac{{m}_c\langle\bar{q}q\rangle}{576\pi^2}\langle\frac{\alpha_{s}{G}{G}}{\pi}\rangle  \int dydz
 \,\left[9+4s\delta\left(s-\overline{m}_c^2\right)\right]\nonumber\\
 && -\frac{{m}_c\langle\bar{q}q\rangle}{576\pi^2}\langle\frac{\alpha_{s}{G}{G}}{\pi}\rangle  \int dy
 \,y \, ,
\end{eqnarray}

\begin{eqnarray}
\rho^V_8(s)&=&-\rho^A_8(s)\, ,
\end{eqnarray}

\begin{eqnarray}
\rho^V_{10}(s)&=&-\rho^A_{10}(s) \, ,
\end{eqnarray}
here $y_{f}=\frac{1+\sqrt{1-4m_c^2/s}}{2}$,
$y_{i}=\frac{1-\sqrt{1-4m_c^2/s}}{2}$, $z_{i}=\frac{y
m_c^2}{y s -m_c^2}$, $\overline{m}_c^2=\frac{(y+z)m_c^2}{yz}$,
$ \widetilde{m}_c^2=\frac{m_c^2}{y(1-y)}$, $\int dydz=\int_{y_i}^{y_f}dy \int_{z_i}^{1-y}dz $ and $\int dy=\int_{y_i}^{y_f}dy $.
 When the $\delta$ functions $\delta\left(s-\overline{m}_c^2\right)$ and $\delta\left(s-\widetilde{m}_c^2\right)$ appear,
 $\int_{y_i}^{y_f}dy \to \int_{0}^{1}dy$, $\int_{z_i}^{1-y}dz \to \int_{0}^{1-y}dz$.

 From the analytical expressions of the spectral densities $\rho_{10}^{A/V}(s)$ involving  the vacuum condensates of dimension $10$, the highest dimensional vacuum condensates, we can see explicitly that the $\langle\bar{q}g_{s}\sigma Gq\rangle^2$ and $ \langle\bar{q}q\rangle^2 \langle\frac{\alpha_{s}GG}{\pi}\rangle$ are companied with the inverted Borel parameters $\frac{1}{T^2}$, $\frac{1}{T^4}$ or $\frac{1}{T^6}$, their contributions are greatly amplified at the small values of the $T^2$, and they play a great important role in determining the lowest values of the Borel parameters. In the spectral densities $\rho_{8}^{A/V}(s)$, some terms involving  the  vacuum condensate $\langle\bar{q}q\rangle\langle\bar{q}g_{s}\sigma Gq\rangle$ are companied with the inverted Borel parameter $\frac{1}{T^2}$ besides the small  numerical denominators, and their contributions are also amplified at the small values of the $T^2$ and also play a great important role in determining the lowest values of the Borel parameters.

   In the spectral density $\rho_{7}^{A}(s)$, there exists a term companied with the  inverted Borel parameter $\frac{1}{T^2}$, while in the spectral density $\rho_{7}^{V}(s)$, there exists no term companied with the  inverted Borel parameter $\frac{1}{T^2}$ at all. Even in the case of  small Borel parameters,
 the contributions of the vacuum condensate $ \langle\bar{q}q\rangle \langle\frac{\alpha_{s}GG}{\pi}\rangle$ are very small in all the QCD sum rules for the hidden-charm tetraquark (molecular) states  due to the small value of the $ \langle\bar{q}q\rangle \langle\frac{\alpha_{s}GG}{\pi}\rangle$ and the associated large numerical denominators \cite{WZG-CTP-DvDvDv,WZG-HT-PRD,WZG-PRD-hidden-charm,Wang-tetra-formula,WZG-4260-4360-4660}, the vacuum condensate $ \langle\bar{q}q\rangle \langle\frac{\alpha_{s}GG}{\pi}\rangle$   plays  a tiny role both in   and  out the Borel windows. While in the spectral densities $\rho_{6}^{A/V}(s)$,
the contributions of the vacuum condensate $ \langle\bar{q}q\rangle^2$ are greatly enhanced due to the associated small numerical denominators. For the conventional mesons, the vacuum condensate $ \langle\bar{q}q\rangle^2$ is always companied with the strong fine-structure constant $\alpha_s(\mu)$ and its contribution is greatly suppressed, and the conventional mesons are not the ideal channels to explore the non-factorizable effects \cite{Review-kappa-kappa}.

 All in all, the higher dimensional vacuum condensates $\langle\bar{q}q\rangle^2$,
 $\langle\bar{q}q\rangle\langle\bar{q}g_{s}\sigma Gq\rangle$,
 $\langle\bar{q}g_{s}\sigma Gq\rangle^2$ and $ \langle\bar{q}q\rangle^2 \langle\frac{\alpha_{s}GG}{\pi}\rangle$ play an important role in determining the Borel windows, their positive or negative contributions, say $\rho^V_6(s)=-\rho^A_6(s)$, $\rho^V_8(s)=-\rho^A_8(s)$ and $\rho^V_{10}(s)=-\rho^A_{10}(s)$, have great influences
 on the convergent behaviors of the operator product expansion and the pole contributions. We can take account of the non-factorizable  effects by introducing a parameter $\kappa$,
 \begin{eqnarray}
 \langle\bar{q}q\rangle^2                                         &\to & \kappa\, \langle\bar{q}q\rangle^2\, , \nonumber\\
  \langle\bar{q}q\rangle\langle\bar{q}g_{s}\sigma Gq\rangle       &\to & \kappa\, \langle\bar{q}q\rangle\langle\bar{q}g_{s}\sigma Gq\rangle\, , \nonumber\\
  \langle\bar{q}g_{s}\sigma Gq\rangle^2                           &\to & \kappa\, \langle\bar{q}g_{s}\sigma Gq\rangle^2\, , \nonumber\\
  \langle\bar{q}q\rangle^2 \langle\frac{\alpha_{s}GG}{\pi}\rangle &\to & \kappa\, \langle\bar{q}q\rangle^2 \langle\frac{\alpha_{s}GG}{\pi}\rangle\, .
 \end{eqnarray}
We should bear in mind that the parameter $\kappa$ associated with the vacuum condensates of dimensions $6$, $8$ and $10$ is not necessary to have the same value. The four-quark vacuum condensate $\langle\bar{q}q\rangle^2$ serves  as a milestone  for the hidden-charm (or hidden-bottom) tetraquark (molecular) states \cite{WZG-CTP-DvDvDv,WZG-HT-PRD,WZG-PRD-hidden-charm,Wang-tetra-formula,WZG-4260-4360-4660}, the dominant influence of the higher dimensional vacuum condensates  comes from the vacuum condensate $\langle\bar{q}q\rangle^2$ near the Borel windows, the Borel windows are sensitive to the coefficient $\kappa$ associated with the  $\langle\bar{q}q\rangle^2$.

In the present work, we take account of the non-factorizable effects of the vacuum condensates $\langle\bar{q}q\rangle\langle\bar{q}g_{s}\sigma Gq\rangle$,
$\langle\bar{q}g_{s}\sigma Gq\rangle^2$ and $\langle\bar{q}q\rangle^2 \langle\frac{\alpha_{s}GG}{\pi}\rangle$ for a more robust estimation, as they are also acquired via the vacuum saturation after all, and we will simplify the analysis and set all the parameters  $\kappa$  to have the same values.

 We  differentiate  Eq.\eqref{QCDSR} with respect   to  $\tau=\frac{1}{T^2}$, and eliminate the
 pole residues  $\lambda_{Z}$ to obtain the QCD sum rules for the tetraquark masses,
 \begin{eqnarray}
 M^2_{Z}= \frac{-\frac{d}{d \tau }\int_{4m_c^2}^{s_0} ds\,\rho_{QCD}(s)\,e^{-\tau s}}{\int_{4m_c^2}^{s_0} ds \, \rho_{QCD}(s)\, e^{-\tau s}}\, .
\end{eqnarray}

\section{Numerical results and discussions}
We adopt two sets of input parameters at the QCD side to explore  the hidden-charm tetraquark states. \\ \\
{\bf Set I}

 We choose the standard values  of  the  vacuum condensates
$\langle\bar{q}q \rangle=-(0.24\pm 0.01\, \rm{GeV})^3$,  $\langle\bar{q}g_s\sigma G q \rangle=m_0^2\langle \bar{q}q \rangle$,
 $m_0^2=(0.8 \pm 0.1)\,\rm{GeV}^2$, $\langle \frac{\alpha_s
GG}{\pi}\rangle=0.012\pm0.004\,\rm{GeV}^4$    at the  energy scale  $\mu=1\, \rm{GeV}$
\cite{SVZ79-1,SVZ79-2,PRT85,Ioffe-NPB-1981,Ioffe-mixcondensate,ColangeloReview}, and  take the $\overline{MS}$ mass of the charm  quark from the Particle Data Group, $m_{c}(m_c)=(1.275\pm0.025)\,\rm{GeV}$ \cite{PDG}.
In addition,  we take  account of the energy-scale dependence of  all the input parameters \cite{Narison-mix},
 \begin{eqnarray}
 \langle\bar{q}q \rangle(\mu)&=&\langle\bar{q}q\rangle({\rm 1 GeV})\left[\frac{\alpha_{s}({\rm 1 GeV})}{\alpha_{s}(\mu)}\right]^{\frac{12}{33-2n_f}}\, , \nonumber\\
  \langle\bar{q}g_s \sigma Gq \rangle(\mu)&=&\langle\bar{q}g_s \sigma Gq \rangle({\rm 1 GeV})\left[\frac{\alpha_{s}({\rm 1 GeV})}{\alpha_{s}(\mu)}\right]^{\frac{2}{33-2n_f}}\, ,\nonumber\\
 m_c(\mu)&=&m_c(m_c)\left[\frac{\alpha_{s}(\mu)}{\alpha_{s}(m_c)}\right]^{\frac{12}{33-2n_f}} \, ,\nonumber\\
\alpha_s(\mu)&=&\frac{1}{b_0t}\left[1-\frac{b_1}{b_0^2}\frac{\log t}{t} +\frac{b_1^2(\log^2{t}-\log{t}-1)+b_0b_2}{b_0^4t^2}\right]\, ,
\end{eqnarray}
  where $t=\log \frac{\mu^2}{\Lambda^2}$, $b_0=\frac{33-2n_f}{12\pi}$, $b_1=\frac{153-19n_f}{24\pi^2}$, $b_2=\frac{2857-\frac{5033}{9}n_f+\frac{325}{27}n_f^2}{128\pi^3}$,  $\Lambda=213\,\rm{MeV}$, $296\,\rm{MeV}$  and  $339\,\rm{MeV}$ for the quark flavor numbers  $n_f=5$, $4$ and $3$, respectively  \cite{PDG}.
In the present work, we explore  the hidden-charm tetraquark   states  and choose $n_f=4$, then  evolve all the
input parameters  to some typical energy scales $\mu$ to extract the tetraquark  masses. The value $\langle \frac{\alpha_s
GG}{\pi}\rangle=0.012\pm0.004\,\rm{GeV}^4$ is still frequently used  in the current QCD sum rules analysis, as no significant progress in its determination
has ever been made and proven   to be more reliable. \\ \\
{\bf Set II}

We choose the updated parameters obtained by S. Narison, including the mass of the charm quark and the gluon condensate, $m_{c}(m_c)=(1.266\pm0.006)\,\rm{GeV}$ and $\langle \frac{\alpha_s
GG}{\pi}\rangle=0.021\pm0.001\,\rm{GeV}^4$ \cite{Narison-2101}, while  the energy scales of the vacuum condensates are set to be  $\mu=1\,\rm{GeV}$.

For the parameter  Set I, we  take the energy scale formula $\mu=\sqrt{M^2_{X/Y/Z}-(2{\mathbb{M}}_c)^2}$ with the effective $c$-quark mass, ${\mathbb{M}}_c=1.82\,\rm{GeV}$, to obtain the suitable energy scales of the QCD spectral densities \cite{Wang-tetra-formula,WZG-EPJC-1.82}.
We can rewrite the energy scale formula in the form,
\begin{eqnarray}\label{formula-Regge}
M^2_{X/Y/Z}&=&\mu^2+{\rm Constants}\, ,
\end{eqnarray}
where the Constants have the value $4{\mathbb{M}}_c^2$ and are fitted by the QCD sum rules, the predicted tetraquark  masses and the pertinent  energy scales of the QCD spectral densities have a  Regge-trajectory-like relation \cite{WZG-CTP-DvDvDv}.

We usually consult the experimental data on  the mass gaps between the ground states and first radial excited states  to  adjust the continuum threshold parameters $s_0$ to exclude the possible contaminations from  the continuum states. Although the tetraquark states have not been experimentally established yet, there are several excellent candidates, such as the $Z_c(3900)$, $Z_c(4020)$, $Z_c(4430)$, etc.

According to the possible quantum numbers, decay modes and mass  gaps, in the scenarios of tetraquark states, we can tentatively assign the $X(3915)$ and $X(4500)$ as the 1S and 2S  hidden-charm  tetraquark states with the $J^{PC}=0^{++}$ \cite{X4140-tetraquark-Lebed,X3915-X4500-EPJC-WZG}, assign
the $Z_c(3900)$ and $Z_c(4430)$   as  the 1S and 2S hidden-charm tetraquark states with the $J^{PC}=1^{+-}$, respectively \cite{Maiani-Z4430-1405,Nielsen-1401,WangZG-Z4430-CTP},   assign the $Z_c(4020)$ and $Z_c(4600)$ as the 1S and 2S hidden-charm tetraquark states with the $J^{PC}=1^{+-}$, respectively  \cite{ChenHX-Z4600-A,WangZG-axial-Z4600}, and assign the $X(4140)$ and $X(4685)$ the 1S and 2S hidden-charm tetraquark states with the $J^{PC}=1^{++}$, respectively \cite{WZG-Di-X4140-EPJC,WZG-X4140-X4685}. The energy gaps between the ground states (1S) and first radial excited states (2S) are about $0.57\sim 0.59 \,\rm{GeV}$. In the present work, we can take the continuum threshold parameters as $\sqrt{s_0}=M_Z+0.4\sim0.6\,\rm{GeV}$, where the $M_Z$ are the ground state  masses of the hidden-charm tetraquark states.

 Direct calculations based on the QCD sum rules indicate that the $C\gamma_5\otimes \gamma_\mu C\pm C\gamma_\mu\otimes \gamma_5 C$ type hidden-charm tetraquark states $c\bar{c}u\bar{d}$ with the $J^{PC}=1^{++}$ and $1^{+-}$ respectively have almost degenerated  masses \cite{WZG-HT-PRD,WZG-PRD-hidden-charm}, while the $C\otimes \gamma_\mu C\pm C\gamma_\mu\otimes  C$ type hidden-charm tetraquark states $c\bar{c}u\bar{d}$ with the  $J^{PC}=1^{-+}$ and $1^{--}$ respectively have
  almost degenerated  masses or slightly different masses
\cite{Wang-tetra-formula,WZG-4260-4360-4660}. In the present work, we do not distinguish the charge conjugation, and take it for granted that the $C\gamma_5\otimes \gamma_\mu C$ type  tetraquark state has a mass about $3.9\,\rm{GeV}$, just like the $Z_c(3900)$, and tentatively choose the continuum threshold parameter as $\sqrt{s_0}=4.4\,\rm{GeV}$ as a guide to  optimize the parameters.

In 2019, the LHCb collaboration  explored  the $m(J/\psi \pi^-)$ versus $m(K^+\pi^-)$ plane in the decays $B^0\to J/\psi K^+\pi^-$, and observed two possible  structures near $m(J/\psi \pi^-)=4200 \,\rm{MeV}$ and $4600\,\rm{MeV}$, respectively \cite{LHCb-Z4600}.  The structure near $m(J/\psi \pi^-)=4600 \,\rm{MeV}$ is in very good  agreement with our predicted  mass of the $C\otimes \gamma_\mu C- C\gamma_\mu\otimes  C$  type tetraquark state with the  $J^{PC}=1^{--}$ \cite{WZG-4260-4360-4660}.
In Ref.\cite{WZG-Vector-4600}, we assign the $Z_c(4600)$ to be the 1S vector tetraquark state tentatively  and study its two-body strong decays with the QCD sum rules based on solid quark-hadron duality, and obtain the total decay width $144.8^{+50.6}_{-33.9}\,{\rm{MeV}}$, which is reasonable for the tetraquark state.

In the scenario of tetraquark states, there are two possible assignments for the $Z_c(4600)$, one is the first radial excited state of the $Z_c(4020)$ with the $J^{PC}=1^{+-}$ \cite{ChenHX-Z4600-A,WangZG-axial-Z4600}, the other is the 1S vector tetraquark state  with the $J^{PC}=1^{--}$ \cite{WZG-Vector-4600},  more experimental data are still needed to obtain a more reliable assignment. At the present time,  we take it for granted that the $C\otimes \gamma_\mu C$ type tetraquark state has a mass about $4.6\,\rm{GeV}$,  and choose the continuum threshold parameter as $\sqrt{s_0}=5.1\,\rm{GeV}$ as a guide to  optimize the parameters.

\begin{figure}
\centering
\includegraphics[totalheight=6cm,width=7cm]{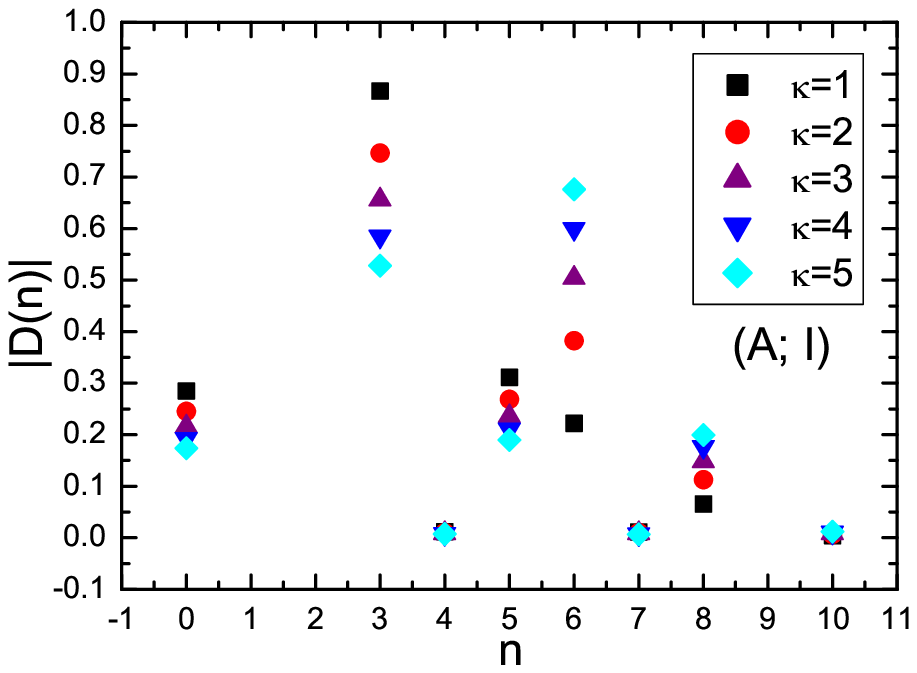}
\includegraphics[totalheight=6cm,width=7cm]{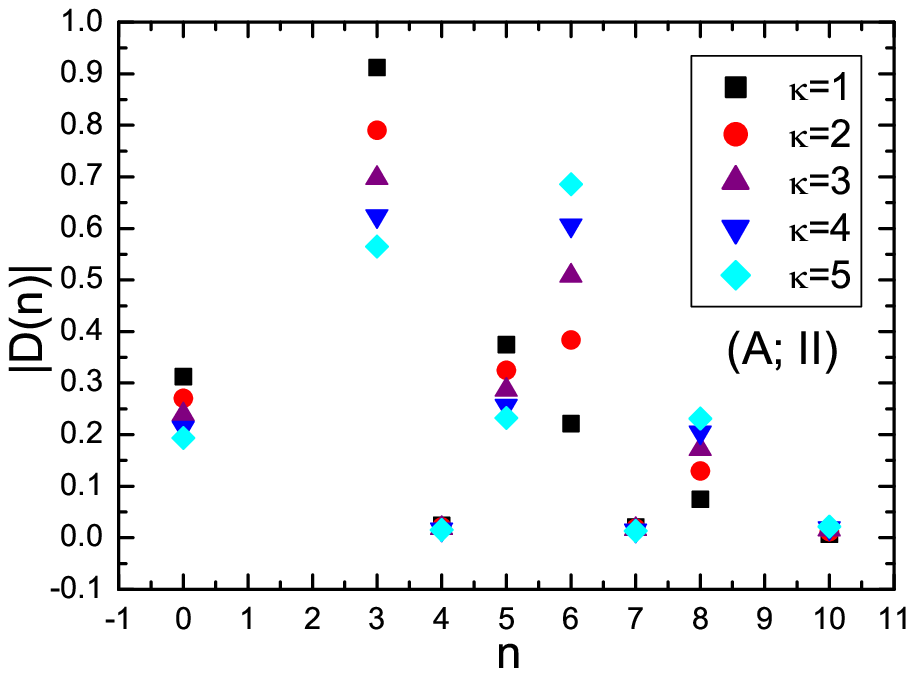}
\includegraphics[totalheight=6cm,width=7cm]{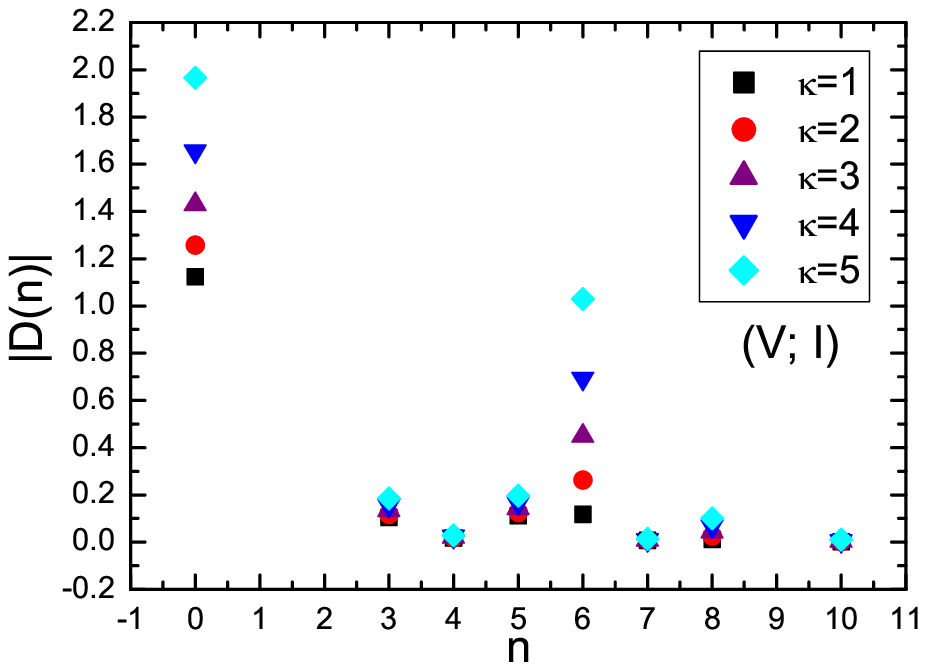}
\includegraphics[totalheight=6cm,width=7cm]{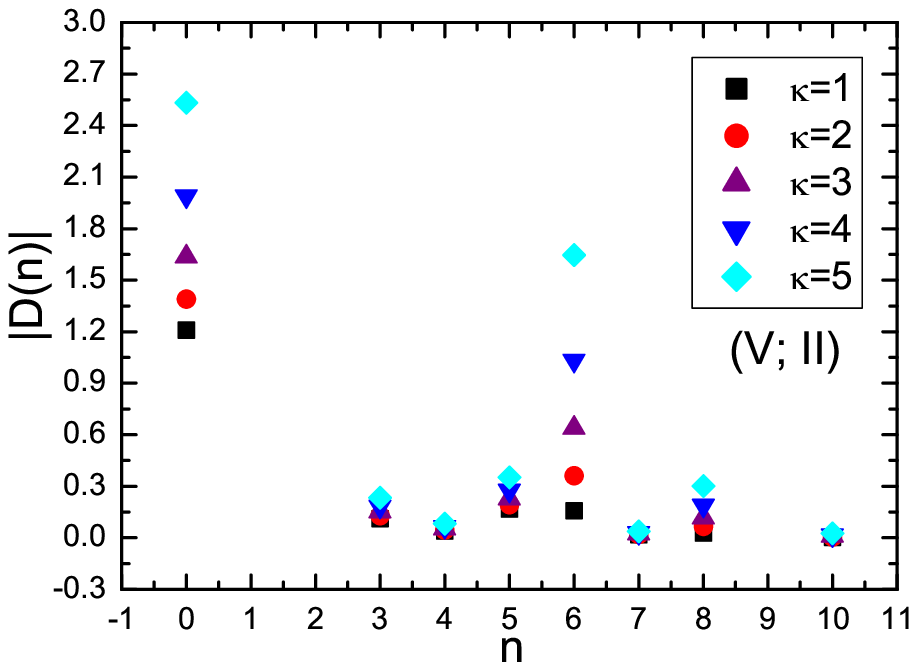}
  \caption{ The absolute contributions of the vacuum condensates, where the A and V denote the axialvector and vector tetraquark states, respectively, the I and II denote the parameters Set I and Set II, respectively.   }\label{OPE-AV}
\end{figure}
\begin{figure}
\centering
\includegraphics[totalheight=6cm,width=7cm]{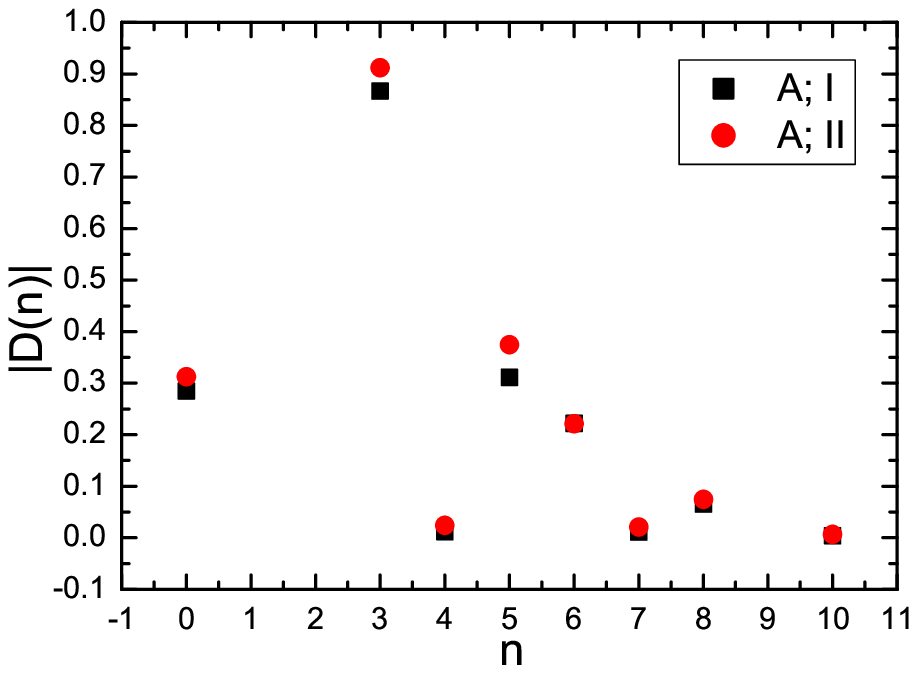}
\includegraphics[totalheight=6cm,width=7cm]{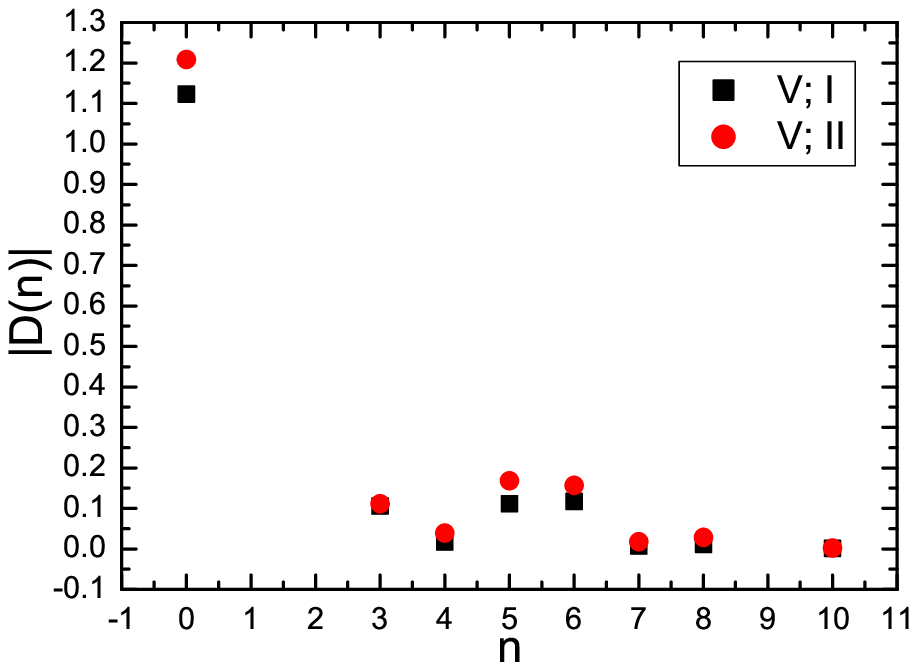}
  \caption{ The absolute contributions of the vacuum condensates for $\kappa=1$, where the A and V denote the axialvector and vector tetraquark states, respectively, the I and II denote the parameters Set I and Set II, respectively.  }\label{OPE-AV-k1}
\end{figure}

In Fig.\ref{OPE-AV}, we plot the absolute values of the contributions of the vacuum condensates in the operator product expansion with the parameters ($\sqrt{s_0}=4.4\,\rm{GeV}$, $T^2=2.9\,\rm{GeV}^2$) and ($5.1\,\rm{GeV}$, $3.9\,\rm{GeV}^2$) for the axialvector and vector hidden-charm tetraquark states, respectively.  In both cases, the contributions of the vacuum condensates of dimension $6$ serve as a milestone. Although the operator product expansion converges for all the values $\kappa=1$, $2$, $\cdots$, $5$, the convergent behavior becomes worse with increase of the values  of the $\kappa$. A good convergent behavior of the operator product expansion requires a small value of the $\kappa$.
In Fig.\ref{OPE-AV-k1}, we plot the absolute values of the contributions of the vacuum condensates in the operator product expansion with the value $\kappa=1$ as an example to examine the influences  of the parameters Set I and Set II. From the figure, we can see  that the parameter Set I leads to a slightly better convergent behavior compared to the parameter Set II.

In the QCD sum rules for the hidden-charm (or hidden-bottom) tetraquark (molecular) states, we adopt the spectral densities $\rho_{QCD}(s)\Theta(s-s_0)$ to represent the contributions of the higher resonances and continuum states, where $\Theta(s)=1$ for $s>0$, else $\Theta(s)=0$ \cite{WZG-CTP-DvDvDv,WZG-HT-PRD,WZG-PRD-hidden-charm,Wang-tetra-formula,WZG-4260-4360-4660}. Such approximations cannot ensure the largest contributions come from the perturbative terms, sometimes the contributions of the vacuum condensates of dimensions of $3$, $5$ or $6$ are even larger than the contributions of the perturbative terms, for example, in the present case for the axialvector hidden-charm tetraquark state, $D(3)\gg D(0)$ and $|D(5)|\approx D(0)$. In all the QCD sum rules, the vacuum condensates $\langle\bar{q}q\rangle^2$ with $q=u$, $d$, $s$ serve a milestone in judging the convergent behaviors of the operator product expansion \cite{WZG-CTP-DvDvDv,WZG-HT-PRD,WZG-PRD-hidden-charm,Wang-tetra-formula,WZG-4260-4360-4660}, we require the hierarchies  $|D(6)|>|D(8)|>|D(10)|$ and $|D(10)|\leq 1\%$ or $\ll 1\%$.

\begin{figure}
\centering
\includegraphics[totalheight=6cm,width=7cm]{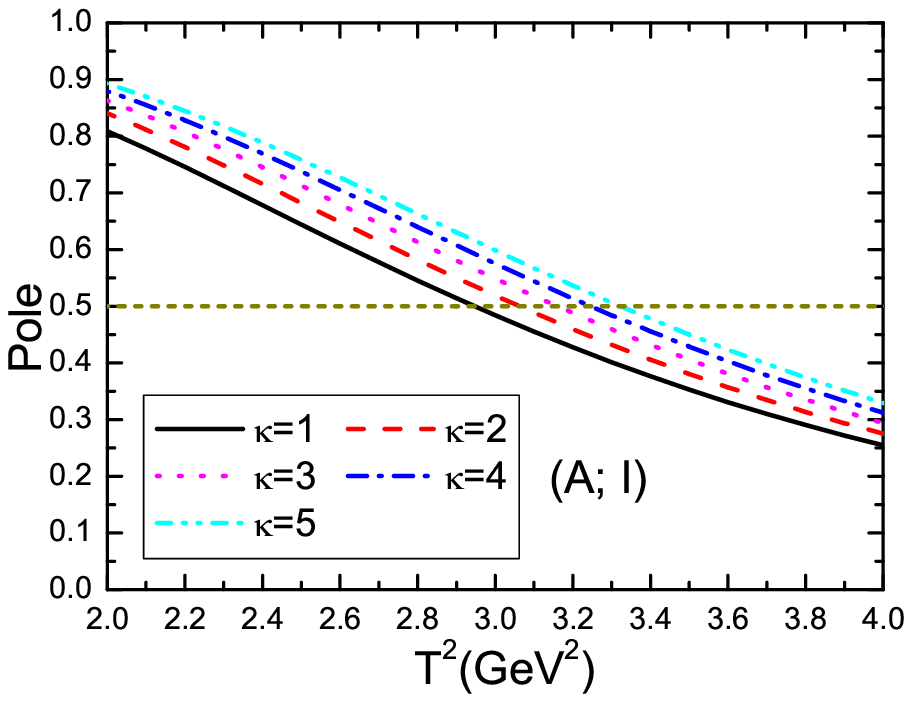}
\includegraphics[totalheight=6cm,width=7cm]{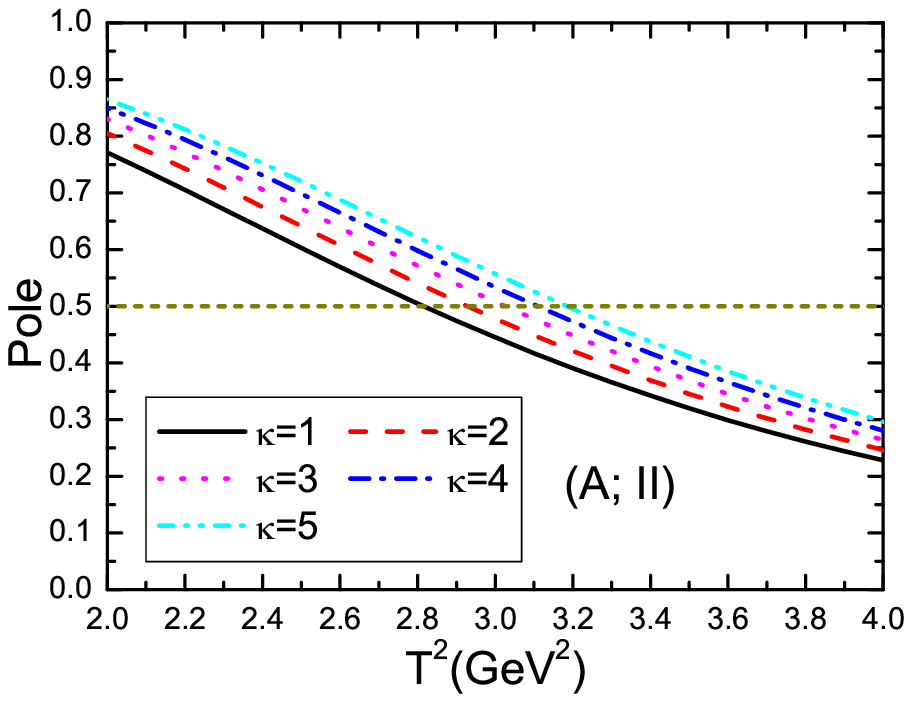}
\includegraphics[totalheight=6cm,width=7cm]{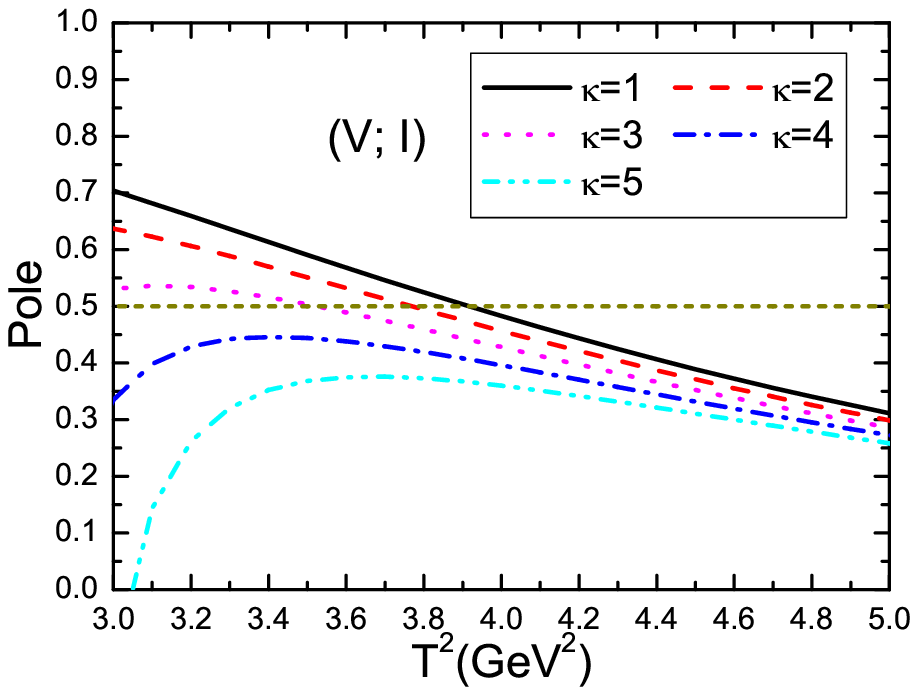}
\includegraphics[totalheight=6cm,width=7cm]{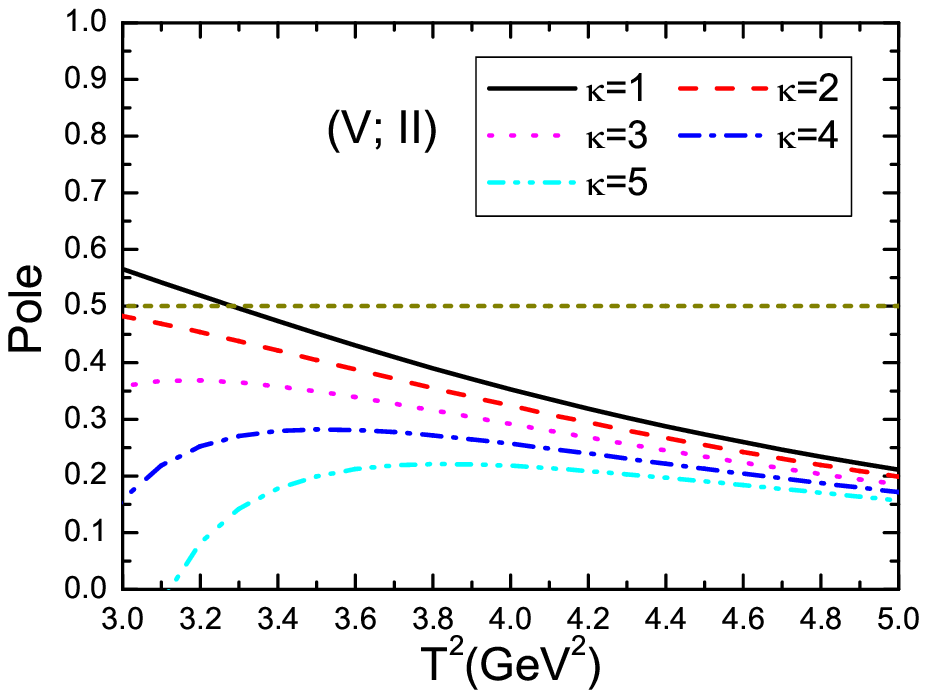}
  \caption{ The pole contributions with variations of the Borel parameters, where the A and V denote the axialvector and vector tetraquark states, respectively, the I and II denote the parameters Set I and Set II, respectively.   }\label{pole-AV}
\end{figure}
\begin{figure}
\centering
\includegraphics[totalheight=6cm,width=7cm]{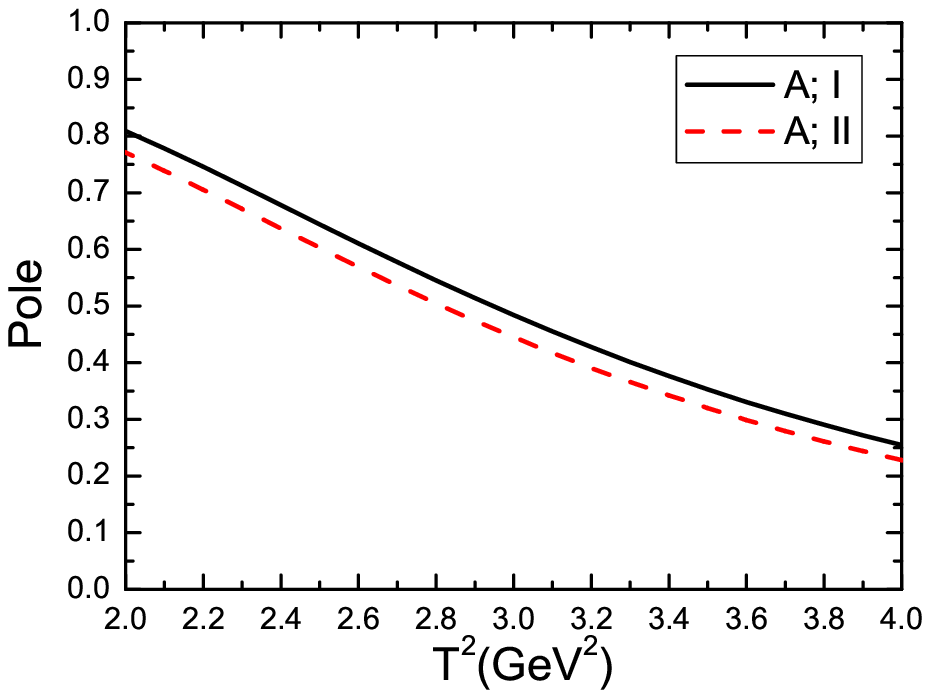}
\includegraphics[totalheight=6cm,width=7cm]{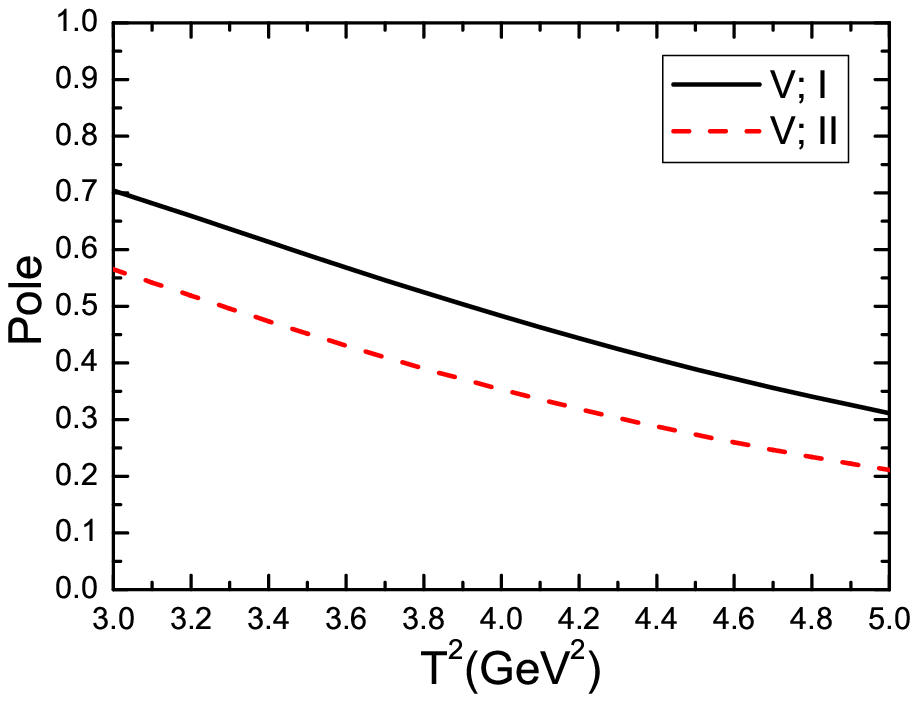}
   \caption{ The pole contributions with variations of the Borel parameters for $\kappa=1$, where the A and V denote the axialvector and vector tetraquark states, respectively, the I and II denote the parameters Set I and Set II, respectively.   }\label{pole-AV-k1}
\end{figure}

In Fig.\ref{pole-AV}, we plot the pole contributions  with the  parameters ($\sqrt{s_0}=4.4\,\rm{GeV}$, $T^2=2.9\,\rm{GeV}^2$) and ($5.1\,\rm{GeV}$, $3.9\,\rm{GeV}^2$) for the axialvector and vector hidden-charm tetraquark states, respectively.  From the figure, we can see  that in the case of the axialvector tetraquark state, the pole contributions increase monotonically with increase of the values of the $\kappa$,  while in the case of the vector tetraquark state, the pole contributions decrease monotonically with increase of the values of the $\kappa$, the pole contributions play an important role in examining the values of the $\kappa$. However, the pole contributions alone cannot optimize the parameter $\kappa$.
In Fig.\ref{pole-AV-k1}, we plot the pole contributions  with the value $\kappa=1$ as an example to examine the influences  of the parameters Set I and Set II. From the figure, we can see  that the parameter Set I leads to a slightly (much) larger pole contribution compared to the parameter Set II for the axialvector (vector) tetraquark state, the pole contributions in the vector tetraquark channels also  play an important role in examining the parameters Set I and Set II.

\begin{figure}
\centering
\includegraphics[totalheight=6cm,width=7cm]{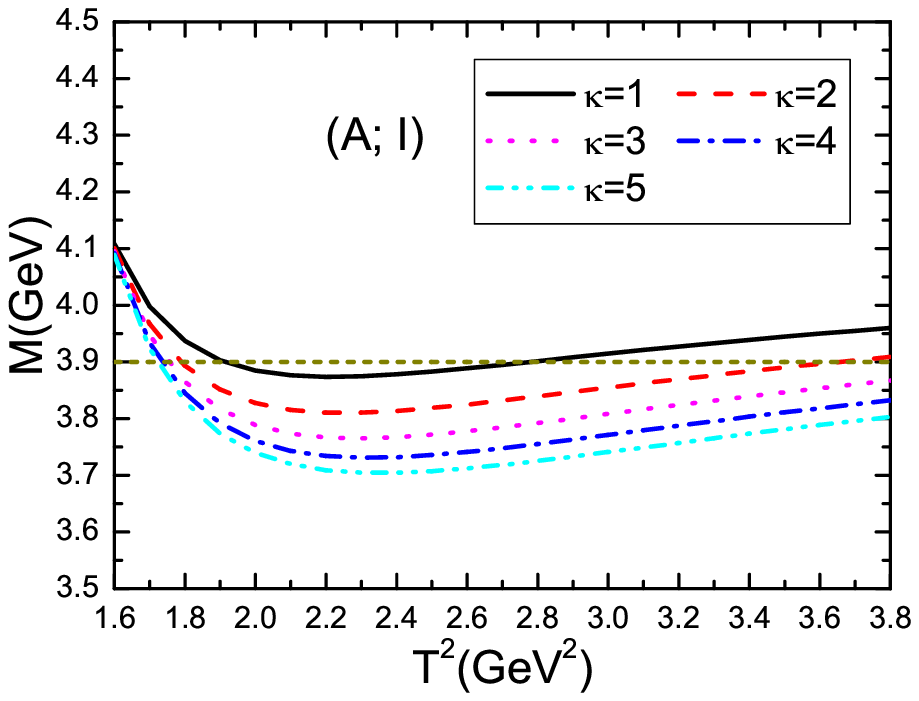}
\includegraphics[totalheight=6cm,width=7cm]{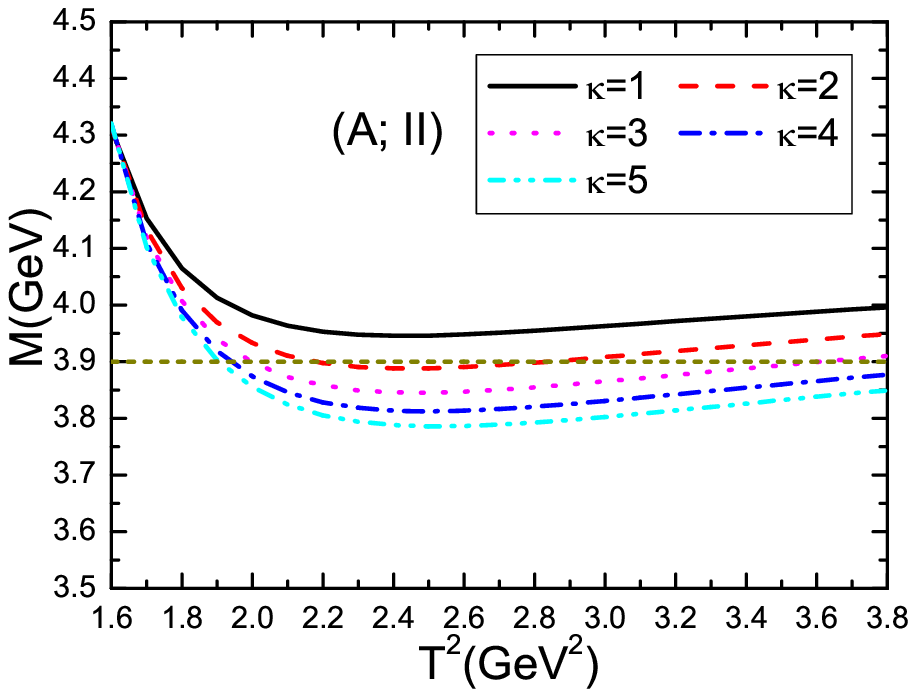}
\includegraphics[totalheight=6cm,width=7cm]{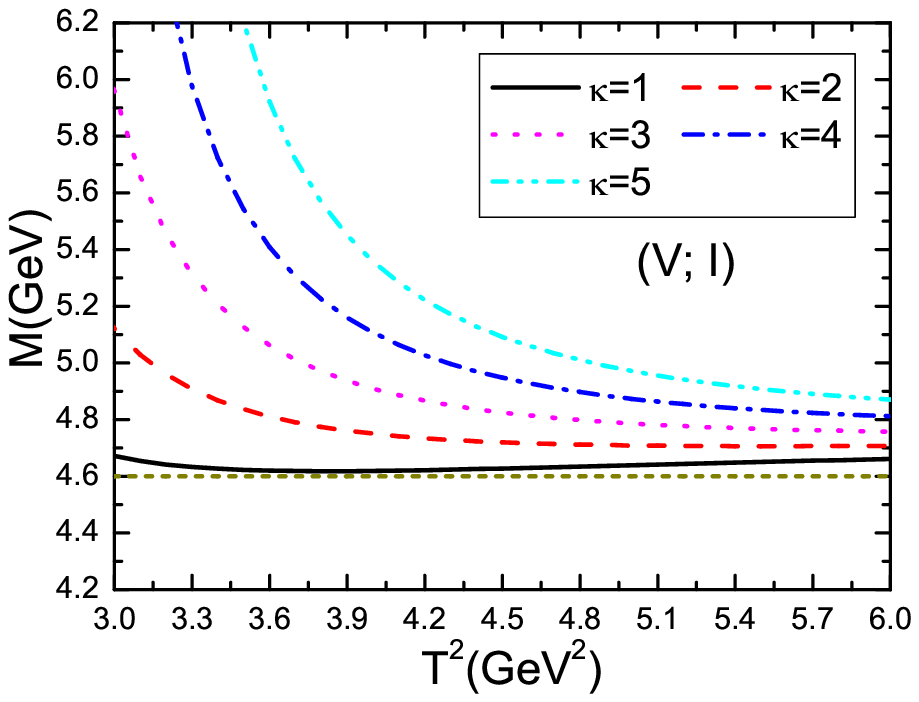}
\includegraphics[totalheight=6cm,width=7cm]{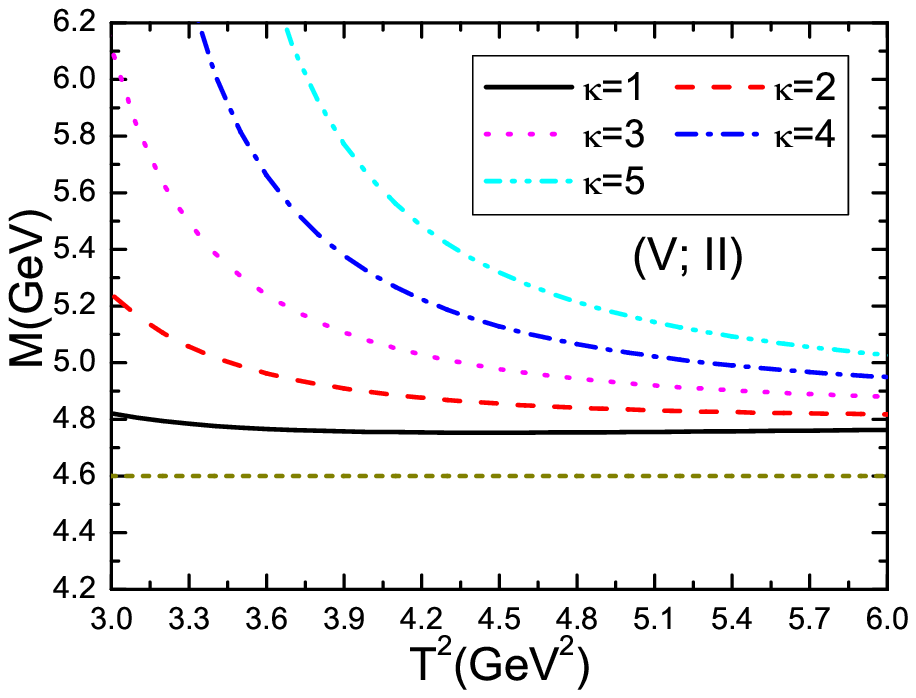}
 \caption{ The tetraquark masses with variations of the Borel parameters, where the A and V denote the axialvector and vector tetraquark states, respectively, the I and II denote the parameters Set I and Set II, respectively.   }\label{mass-AV}
\end{figure}

In Fig.\ref{mass-AV}, we plot the predicted masses via the Borel parameters with the continuum threshold parameters $\sqrt{s_0}=4.4\,\rm{GeV}$ and $5.1\,\rm{GeV}$ for the axialvector and vector hidden-charm tetraquark states, respectively.  From the figure, we can see  that in the case of the axialvector tetraquark state, the predicted masses decrease monotonically and slowly with increase of the values of the $\kappa$,  while in the case of the vector tetraquark state, the predicted masses increase monotonically and quickly with increase of the values of the $\kappa$, furthermore, in the case of the vector tetraquark state, we cannot obtain flat Borel  platforms with increase of the values of the $\kappa$.  We can obtain the conclusion tentatively  that the vector channel is much better to   determine the value of the $\kappa$ compared to the axialvector channel.
Moreover, from the Fig.\ref{mass-AV}, we can see  that the parameter Set II leads to a slightly (much) larger predicted masses compared to the parameter Set I for the axialvector (vector) tetraquark state, the predicted masses in the vector channel also play an important role in examining the parameters Set I and Set II.

According to above analysis, we search for the best Borel parameters $T^2$ and continuum threshold parameters $s_0$ via trial and error to satisfy the two fundamental criteria of the QCD sum rules for the typical values $\kappa=1$, $2$, $3$, $4$ and $5$. The resulting  Borel parameters, continuum threshold parameters, energy scales,  pole contributions are shown plainly in Table \ref{Borel-mass-pole}. In calculations, we choose the same pole contributions  $(40-60)\%$ in all the channels and the central values are larger than (or equal) $50\%$, the pole dominance criterion is well satisfied. On the other hand, the contributions of the higher dimensional vacuum condensates satisfy  the relation $|D(6)|\gg |D(8)|\gg |D(10)|$ for all the values of the $\kappa$, the operator product expansion converges very well. As the two fundamental criteria of the QCD sum rules are all satisfied, we expect to make reliable predictions.

\begin{figure}
\centering
\includegraphics[totalheight=6cm,width=7cm]{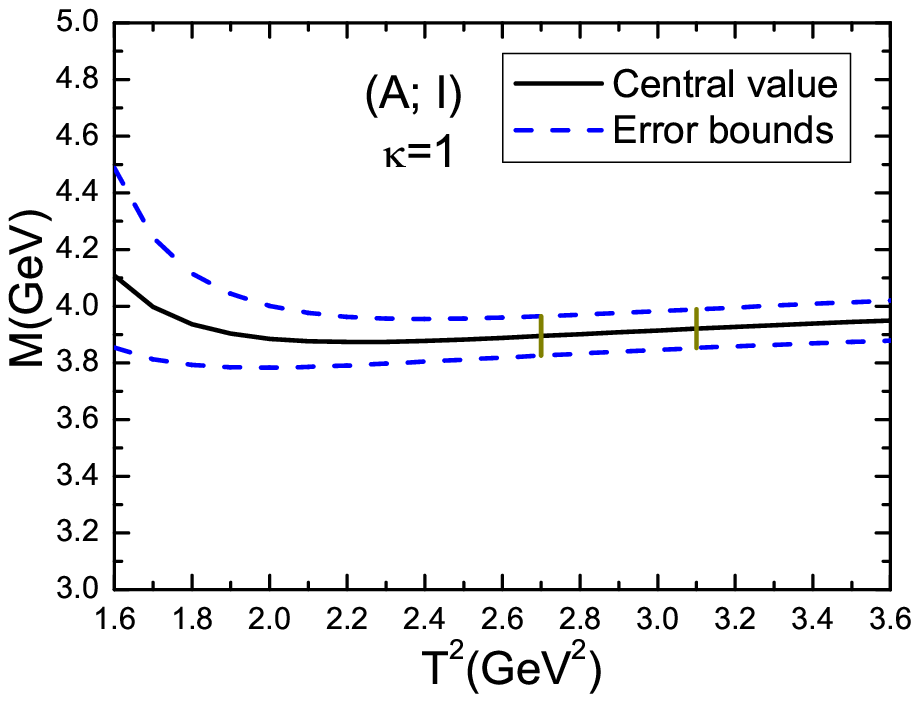}
\includegraphics[totalheight=6cm,width=7cm]{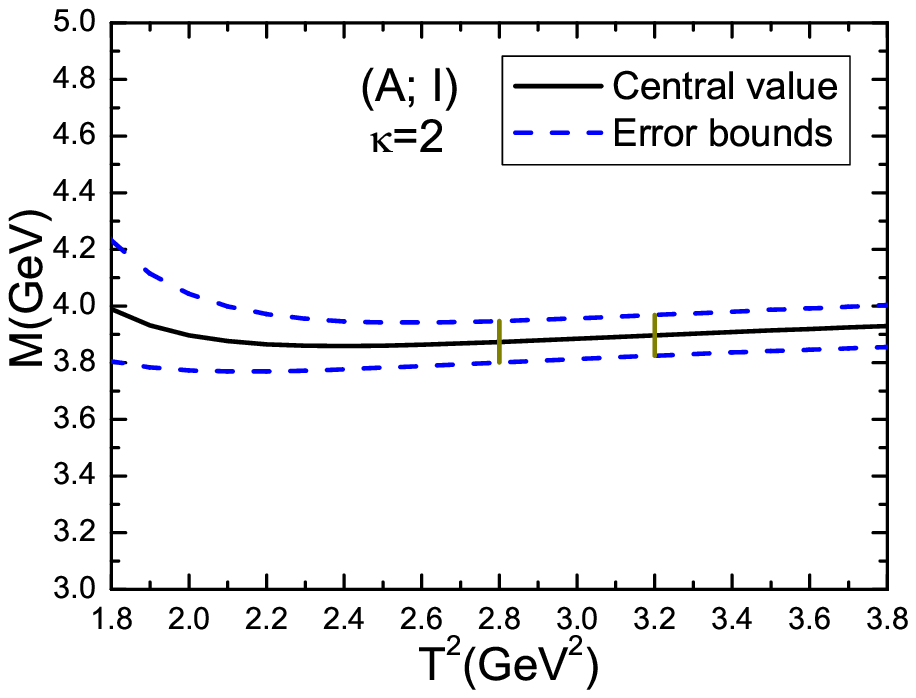}
\includegraphics[totalheight=6cm,width=7cm]{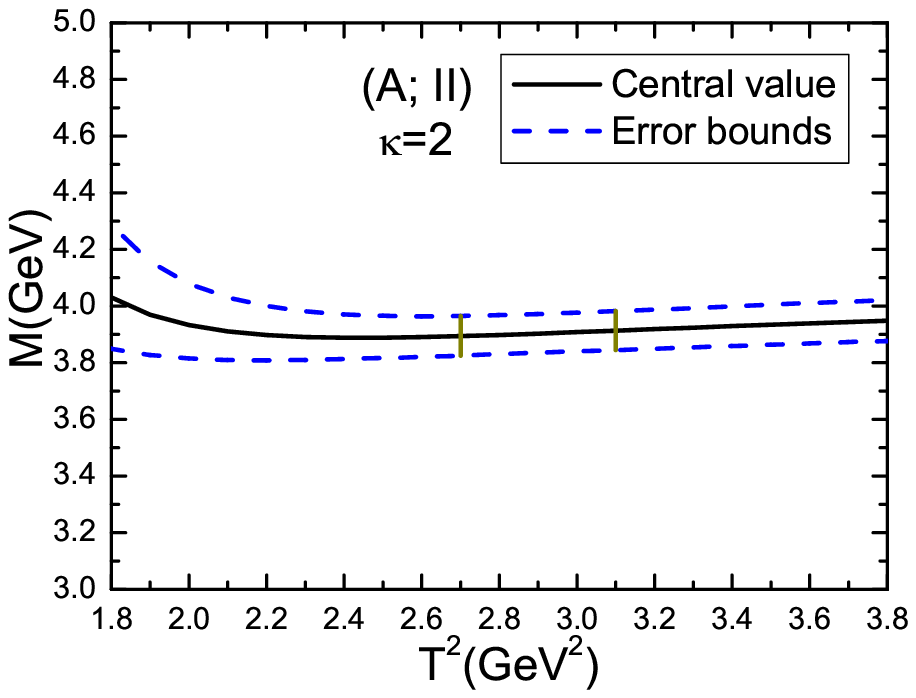}
\includegraphics[totalheight=6cm,width=7cm]{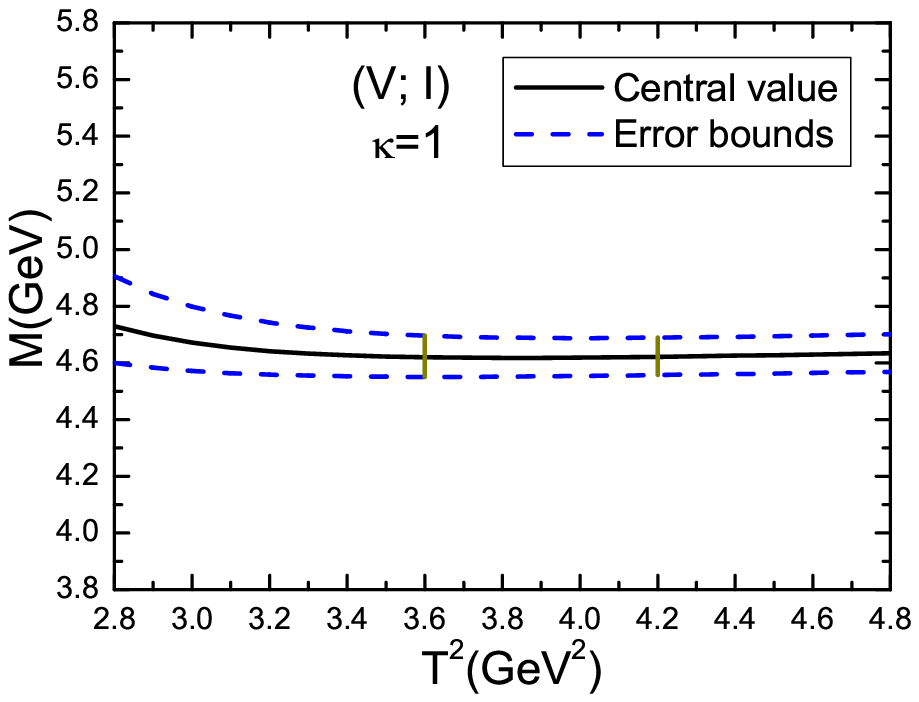}
\includegraphics[totalheight=6cm,width=7cm]{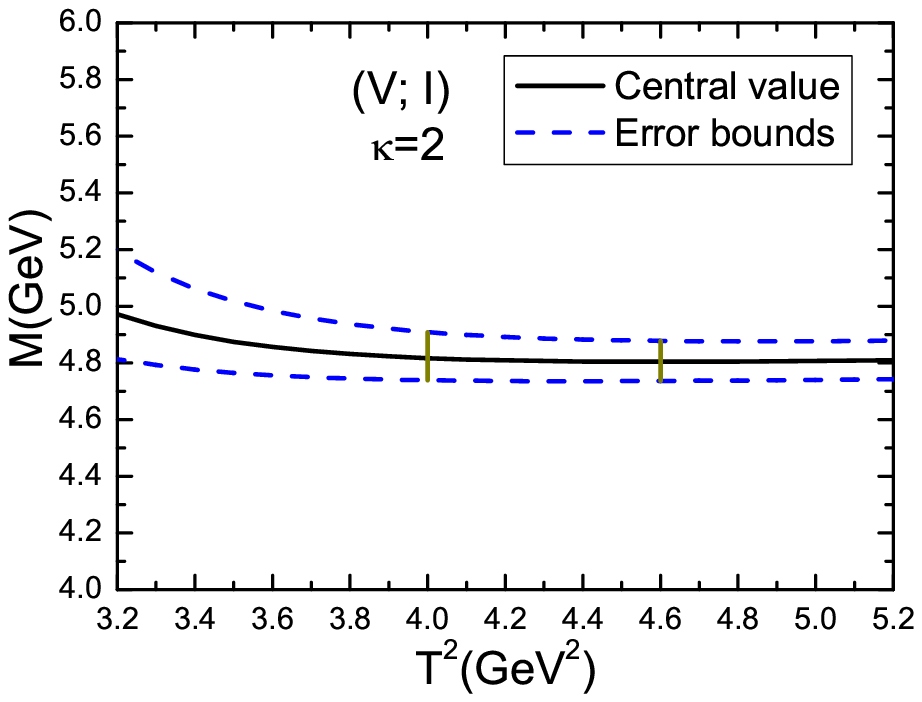}
\includegraphics[totalheight=6cm,width=7cm]{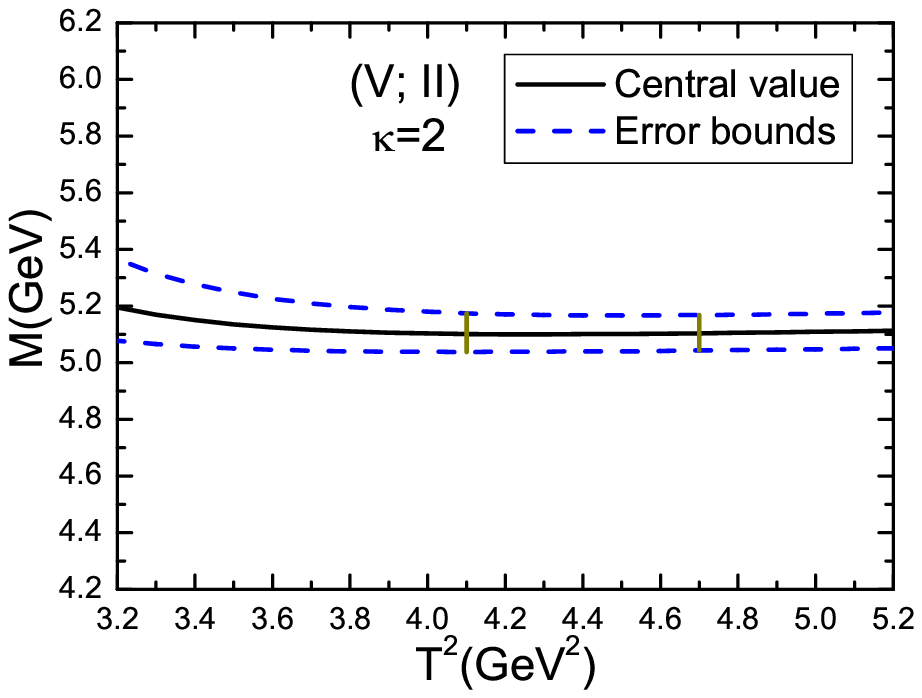}
   \caption{ The tetraquark masses with variations of the Borel parameters, where the A and V denote the axialvector and vector tetraquark states, respectively, the I and II denote the parameters Set I and Set II, respectively.   }\label{mass-AV-UN}
\end{figure}

Now we take account of all uncertainties of the input parameters, and obtain the values of the masses and pole residues of
 the   axialvector and vector hidden-charm tetraquark  states, which are  shown plainly in Fig.\ref{mass-AV-UN} and Table \ref{Borel-mass-pole}.
 In Fig.\ref{mass-AV-UN}, we plot the predicted masses with variations of the Borel parameters at much larger ranges than the Borel windows. From the figure, we can see explicitly  that there appear very flat platforms in the Borel windows, the uncertainties come from the Borel parameters are rather small, the predictions are reasonable and reliable.

\begin{figure}
\centering
\includegraphics[totalheight=8cm,width=10cm]{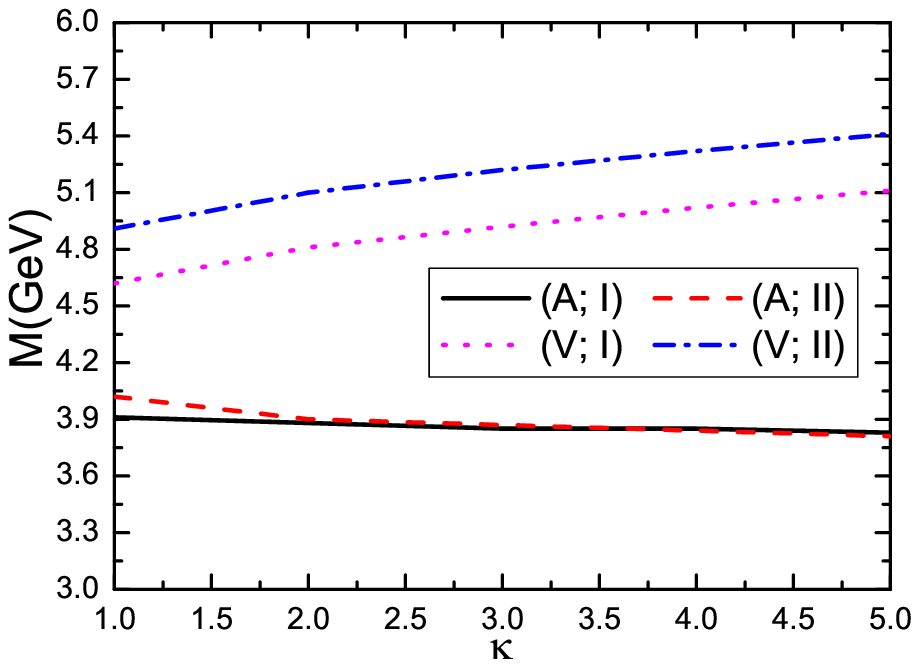}
   \caption{ The tetraquark masses with variations of the  value of the  $\kappa$, where the A and V denote the axialvector and vector tetraquark states, respectively, the I and II denote the parameters Set I and Set II, respectively.   }\label{mass-AV-kappa}
\end{figure}

 In Fig.\ref{mass-AV-kappa}, we plot the central values of the tetraquark masses with variations of the parameter $\kappa$. From Table \ref{Borel-mass-pole} and Fig.\ref{mass-AV-kappa}, we can see  that  for the axialvector tetraquark state, the predicted mass decreases monotonically and slowly with  increase of the parameter $\kappa$, smaller tetraquark mass favors larger value of the $\kappa$. Compared to the parameter Set I, the predicted mass decreases slightly quicker    for the parameter Set II. The axialvector channel is not the best channel to distinguish the parameters Set I and Set II, only by precisely measuring the mass of the axialvector tetraquark state, we can obtain powerful constraint on the input parameters, for example, if the ground state has a mass about $3.9\,\rm{GeV}$, the preferred parameters are (Set I, $\kappa=1$) and  (Set II, $\kappa=2$).

 For the vector tetraquark state, the predicted mass increases monotonically and quickly with increase of the parameter $\kappa$, larger tetraquark mass favors larger value of the $\kappa$, furthermore, the parameter Set II leads to much larger predicted mass compared to the parameter Set I, the vector channel is the best channel to distinguish the input parameters. For example, if the ground state vector tetraquark state has a mass about $4.6\,\rm{GeV}$, which happens to coincide  with the mass of the  $Y(4660)$ or $Z_c(4600)$ (in the isospin limit), the preferred parameters are (Set I, $\kappa=1$). On the other hand, if it has mass about $5.4\,\rm{GeV}$, the preferred parameters  are (Set II, $\kappa=5$).
We can confronted the predictions in Table \ref{Borel-mass-pole} to the experimental data in the future to select the best parameters.

\begin{table}
\begin{center}
\begin{tabular}{|c|c|c|c|c|c|c|c|c|c|}\hline\hline
$J^P$    &$T^2(\rm{GeV}^2)$ &$\sqrt{s_0}(\rm{GeV})$ &$\mu(\rm{GeV})$  &pole         &       &$\kappa$  &$M_{Z}(\rm{GeV})$ &$\lambda_{Z}(10^{-2}\rm{GeV}^5)$\\ \hline

$1^{+}$  &$2.7-3.1$         &$4.4\pm0.1$            &$1.4$            &$(40-63)\%$  &Set I  &$1 $      &$3.91\pm0.08$     &$2.11\pm 0.34$    \\ \hline

$1^{+}$  &$2.8-3.2$         &$4.4\pm0.1$            &$1.3$            &$(39-61)\%$  &Set I  &$2 $      &$3.88\pm0.08$     &$2.01\pm 0.32$    \\ \hline

$1^{+}$  &$2.9-3.3$         &$4.4\pm0.1$            &$1.3$            &$(40-61)\%$  &Set I  &$3 $      &$3.85\pm0.09$     &$2.05\pm 0.31$    \\ \hline

$1^{+}$  &$2.9-3.3$         &$4.4\pm0.1$            &$1.2$            &$(40-61)\%$  &Set I  &$4 $      &$3.85\pm0.09$     &$1.97\pm 0.28$    \\ \hline

$1^{+}$  &$3.0-3.4$         &$4.4\pm0.1$            &$1.2$            &$(40-60)\%$  &Set I  &$5 $      &$3.83\pm0.09$     &$2.02\pm 0.28$    \\ \hline

$1^{+}$  &$2.8-3.2$         &$4.5\pm0.1$            &                 &$(39-61)\%$  &Set II &$1 $      &$4.02\pm0.08$     &$2.24\pm 0.34$    \\ \hline

$1^{+}$  &$2.7-3.1$         &$4.4\pm0.1$            &                 &$(40-62)\%$  &Set II &$2 $      &$3.90\pm0.08$     &$1.95\pm 0.29$    \\ \hline

$1^{+}$  &$2.8-3.2$         &$4.4\pm0.1$            &                 &$(40-62)\%$  &Set II &$3 $      &$3.87\pm0.08$     &$1.98\pm 0.28$    \\ \hline

$1^{+}$  &$2.9-3.3$         &$4.4\pm0.1$            &                 &$(40-61)\%$  &Set II &$4 $      &$3.84\pm0.08$     &$2.01\pm 0.28$    \\ \hline

$1^{+}$  &$3.0-3.4$         &$4.4\pm0.1$            &                 &$(40-60)\%$  &Set II &$5 $      &$3.81\pm0.09$     &$2.05\pm 0.28$    \\ \hline

$1^{-}$  &$3.6-4.2$         &$5.1\pm0.1$            &$2.8$            &$(40-62)\%$  &Set I  &$1 $      &$4.62\pm0.08$     &$6.69\pm 0.81$    \\ \hline

$1^{-}$  &$4.0-4.6$         &$5.3\pm0.1$            &$3.1$            &$(41-61)\%$  &Set I  &$2 $      &$4.81\pm0.09$     &$9.14\pm 1.10$    \\ \hline

$1^{-}$  &$4.3-4.9$         &$5.4\pm0.1$            &$3.3$            &$(40-59)\%$  &Set I  &$3 $      &$4.92\pm0.09$     &$10.8\pm 1.30$    \\ \hline

$1^{-}$  &$4.5-5.1$         &$5.5\pm0.1$            &$3.4$            &$(41-59)\%$  &Set I  &$4 $      &$5.02\pm0.10$     &$12.4\pm 1.50$    \\ \hline

$1^{-}$  &$4.6-5.4$         &$5.6\pm0.1$            &$3.6$            &$(40-61)\%$  &Set I  &$5 $      &$5.11\pm0.10$     &$14.1\pm 1.60$    \\ \hline

$1^{-}$  &$3.7-4.3$         &$5.4\pm0.1$            &                 &$(40-62)\%$  &Set II &$1 $      &$4.91\pm0.07$     &$6.65\pm 0.83$    \\ \hline

$1^{-}$  &$4.1-4.7$         &$5.6\pm0.1$            &                 &$(41-61)\%$  &Set II &$2 $      &$5.10\pm0.07$     &$8.86\pm 0.99$    \\ \hline

$1^{-}$  &$4.3-4.9$         &$5.7\pm0.1$            &                 &$(41-61)\%$  &Set II &$3 $      &$5.22\pm0.08$     &$10.3\pm 1.20$    \\ \hline

$1^{-}$  &$4.5-5.2$         &$5.8\pm0.1$            &                 &$(40-60)\%$  &Set II &$4 $      &$5.32\pm0.09$     &$11.8\pm 1.40$    \\ \hline

$1^{-}$  &$4.6-5.4$         &$5.9\pm0.1$            &                 &$(41-62)\%$  &Set II &$5 $      &$5.41\pm0.09$     &$13.4\pm 1.50$    \\ \hline

\hline
\end{tabular}
\end{center}
\caption{ The Borel parameters, continuum threshold parameters, energy scales,  pole contributions, parameter $\kappa$, masses and pole residues for  the tetraquark  states. }\label{Borel-mass-pole}
\end{table}

\section{Conclusion}
In the QCD sum rules for the tetraquark (molecular) states, the higher dimensional vacuum condensates play an important role in extracting the tetraquark masses.
In the present work, we  construct  the $C\gamma_5\otimes \gamma_\mu C$ type  and  $C\otimes \gamma_\mu C$ type four-quark currents  to investigate the
 tetraquark states  via the QCD sum rules so as to examine the vacuum saturation or factorization approximation for the higher dimensional vacuum condensates.  We carry out the operator product expansion up to  the vacuum condensates of dimension-10  consistently, and introduce a parameter $\kappa$ to parameterize the derivation from the vacuum saturation or factorization approximation, and choose two sets parameters to examine the tetraquark  masses.  In calculations, we observe that smaller  value of the $\kappa$ leads to better convergent behavior in the operator product expansion.

 For the axialvector tetraquark state, the predicted masses decrease monotonically and slowly with  increase of the parameter $\kappa$, larger value of the $\kappa$ leads to smaller tetraquark mass, both the parameters Set I and Set II can lead to a tetraquark mass about $3.9\,\rm{GeV}$, the axialvector channel is not the best channel to distinguish the parameters Set I and Set II. For the vector tetraquark state, the predicted masses increase monotonically and quickly with increase of the parameter $\kappa$, larger value of the $\kappa$ leads to larger tetraquark mass, moreover, the parameter Set II leads to much larger predicted tetraquark mass compared to the parameter Set I, the vector channel is the best channel to distinguish the value of the $\kappa$ and the parameters Set I and Set II.

 If the $Z_c(3900)$ can be assigned  to be the lowest axialvector hidden-charm tetraquark state, the preferred parameters are (Set I, $\kappa=1\sim2$) or (Set II, $\kappa=2\sim3$). On the other hand, if the lowest axialvector hidden-charm tetraquark state has a mass about $3.8\,\rm{GeV}$, the preferred parameters are (Set I or Set II, $\kappa= 5$).
 At the vector sector,  if the $Y(4660)$ can be assigned to be the vector hidden-charm tetraquark state with an implicit P-wave in the (anti)diquark, the preferred parameters are (Set I, $\kappa=1$); else,  if such a vector hidden-charm tetraquark configuration  has a mass about  $4.8\,\rm{GeV}$, the favored parameters are (Set I,  $\kappa=2$).  We can confronted the predictions in Table \ref{Borel-mass-pole} to the experimental data in the future to select the best parameters, or put powerful constraints on the input parameters.

\section*{Acknowledgements}
This  work is supported by National Natural Science Foundation, Grant Number  12175068.

\end{document}